\def\Vec#1{\mbox{\bf $#1$}}
\def\eqne{\end{equation}}
 
\def\eqnb{\begin{equation}}
\def\PTP{Prog. Theor. Phys.(Kyoto)}

\def\NPB{{Nucl. Phys.} {\bf B}}
\def\PLB{{Phys. Lett.} B}
\def\PRL{Phys. Rev. Lett.}
\def\PRD{{Phys. Rev.} D}

\documentclass{PoS}
\usepackage{graphicx}

\newcommand{\Slash}[1]{\ooalign{\hfil/\hfil\crcr$#1$}}
\title{A study of quark-gluon vertices using the lattice Coulomb gauge domain wall fermion}

\ShortTitle{A study of quark-gluon vertices using the lattice Coulomb gauge DWF}

\author{\speaker{Sadataka Furui}\thanks{URL:http://albert.umb.teikyo-u.ac.jp/furui\_lab/furuipbs.htm}\\
        School of Science and Engineering, Teikyo University, Utsunomiya 320-8551, Japan\\
        E-mail: \email{furui@umb.teikyo-u.ac.jp}}

\abstract{I calculate the quark-gluon vertex of the tensor type 
$\Gamma(p,q)=g_3(p,q)p_4 \Slash{\Vec q}$, vector type $g_2(p,q)\Vec q$ and scalar type $g_1(p,q)$, for a small spacial momentum transfer $q=\bf q$ using the gauge configuration of the Domain Wall Fermion (DWF) provided by the RBC-UKQCD collaboration. 

The quark propagator of Coulomb gauge in the cylinder cut, i.e. the four momentum $p$ is directed along the diagonal of the hyper-cubic 
space has small fluctuation and I use this propagator in the evaluation
of the operators by applying the non-perturbative renormalization method.  

The $q$ dependence of the running coupling $\alpha_{s,g_1}(q)$ is compared with $\alpha_I(q)$ measured by the ghost-gluon vertex in Coulomb gauge and $\alpha_s(q)$ measured in Landau gauge.  
}

\FullConference{The XXVI International Symposium on Lattice Field Theory\\
		 July 14-19 2008\\
		 Williamsburg, Virginia, USA}

\begin{document}

\section{Introduction}
In the Lattice2007 symposium we showed \cite{FN07} that the infrared QCD running coupling of ghost-gluon vertices of lattice Coulomb gauge fits better than those of lattice Landau gauge with the experimental running coupling data extracted from the quark-gluon vertices \cite{DBCK06}. The gauge configurations of Kogut-Susskind (KS) fermions of the MILC collaboration \cite{MILC} and those of Domain Wall Fermion (DWF) of RBC/UKQCD collaboration \cite{AABB07} were used in these simulations.

Recently I found that the quark propagator of DWF can be calculated by an extension of the conjugate gradient method that we adopted in the calculation of the
quark propagator of KS fermion or Wilson fermion \cite{SF08}. 
 I calculate the Coulomb gauge quark propagator near the cylinder cut and
small momentum transfer using the DWF gauge configurations and extract the quark-gluon vertex, which is written as
\begin{eqnarray}
\Gamma_\mu(p,q)&=&S^{-1}(p)G_{\mathcal O}(p,q)S^{-1}(p,q)\nonumber\\
&=&\delta^{ab}[g_1(p,q)\gamma_\mu+ig_2(p,q)q_\mu+g_3(p,q)p_\mu\Slash{q}
\end{eqnarray}
as scalar, vector and tensor form factor, respectively. A general method for non-perturbative renormalization of
 lattice operators was presented by Martinelli et al.\cite{MPSTV94} and I adopt
 their method to extract the Coulomb gauge form factors. The lattice simulation of DWF is given in \cite{Chen01}. In the mass-less limit, the wave function possesses exact chiral symmetry and its non-perturbative renormalization  is investigated \cite{Blum02,Blum04}. 

\section{The lattice Coulomb gauge}
 The minimizing function of the Coulomb gauge($\partial_i A_i=0$)
in the Log-$U$ version \cite{FN06} is $F_U[g]=||{\Vec A}^g||^2=\sum_{x,i}{\rm tr}
 \left({{A^g}_{x,i}}^{\dag}A^g_{x,i}\right)$.
Remnant gauge fixing is not done although the gauge field $A_0(x)$ can be further fixed by the following minimizing function via the gauge transformation $g(x_0)$.

The gauge configurations that I adopted are summarized in Table 1.
\begin{table}[htb]
\begin{center}\label{datab}
\begin{tabular}{c c c c c c c c}
   &$\beta$ &N$_f$ &$m$  &$1/a$(GeV)& $L_s$ & $L_t$ &$a L_s$(fm) \\
\hline
DWF$_{01}$ &2.13($\beta_{I}$)&2+1 & 0.01/0.04 & 1.743(20) &16 & 32 & 1.81   \\
DWF$_{02}$ &2.13($\beta_{I}$)&2+1 & 0.02/0.04 & 1.703(16) &16 & 32 & 1.85  \\
DWF$_{03}$ &2.13($\beta_{I}$)&2+1 & 0.03/0.04 & 1.662(20) &16 & 32 & 1.90  \\
\hline
\end{tabular}
\caption{The parameters of the lattice configurations.}
\end{center}
\end{table}
In the process of conjugate gradient iteration, I search the shift parameter 
$\alpha_k^L$ for $\phi_L$ and $\alpha_k^R$ for $\phi_R$ as follows. In the first
50 steps I choose $\alpha_k=Min(\alpha_k^L,\alpha_k^R)$ and shift 
$\phi_{k+1}^L=\phi_k^L-\alpha_k\phi_k^L$ and $\phi_{k+1}^R=\phi_k^R-\alpha_k\phi_k^R$ and in the last 25 steps I choose $\alpha_k=Max(\alpha_k^L,\alpha_k^R)$,
so that the stable solution is selected for both $\phi_L$ and $\phi_R$.

The convergence condition attained in this method is about $0.5\times 10^{-4}$. One can improve the condition by increasing the number of iteration, but the 
overlap of the solution and the plane wave do not change significantly.

In our Lagrangian there is a freedom of choosing global chiral angle in the 5th direction,
\begin{equation}
\psi \to e^{i\eta \gamma_5}\psi, \qquad \bar\psi \to \bar \psi e^{-i\eta\gamma_5}\psi.\nonumber
\end{equation}

On the lattice, the expectation value of the quark propagator $S(p)$ consists of spin dependent $\mathcal{A}p$ part and 
spin independent ${\mathcal B}$ part. 
 I specify the propagator of the left-handed quark by the suffix $L$  and the right-handed quark by the suffix $R$. With use of the renormalization factor $Z$, they are defined as

\[
{\rm Tr} \langle \bar\chi(p,s) P_{L/R} \Psi(p,s)\rangle=Z(p)(2N_c) {\mathcal B}_{L/R}(p,s),
\]
and
\[
{\rm Tr} \langle \bar\chi(p,s) i\Slash{p}P_{L/R}\Psi(p,s)\rangle=Z(p)/(2N_c) i{\bf p}{\mathcal A}_{L/R}(p,s),
\]
where $\displaystyle p_i=\frac{1}{a}\sin \frac{2\pi \bar p_i}{n_i}$ ($\bar p_i=0,1,2,\cdots,n_i/4$). The fifth coordinate of the DWF is specified by $s=0,1,\cdots, 15$.

We parametrize the mass function ${\mathcal M}=\mathcal B/\mathcal A$ as
\begin{equation}
{\mathcal M}(p)=\frac{c\Lambda^{2\alpha+1}}{p^{2\alpha} + \Lambda^{2\alpha}} + \frac{m_f}{a}
\end{equation}
I tried $\alpha=1, 1.25$ and 1.5, compared $\chi^2$ and found that $\alpha=1.25$  gives the best global fit. Since the pole mass $Q^{(w)}$ is not
included in these plots, $m_f$ is set to be 0 here.

In the case of KS fermion of $m_f=0.0136$GeV and 0.027GeV data \cite{FN05,FN06}, we fixed $\alpha=2$ and obtained $\Lambda=0.82$GeV and 0.89GeV, respectively. In general $\Lambda$ becomes larger for larger $\alpha$, but $\Lambda$ of DWF seems larger than that of KS fermion.  In the case of KS fermion, $\Lambda$ becomes
smaller for smaller mass $m_f$, but in the case of DWF, it is opposite.

\begin{table}
\begin{center}
\begin{tabular}{cccccc}
 &$m_{ud}/a$ & $m_s/a$ & $c$ &  $\Lambda$(GeV) & $\alpha$ \\
\hline
DWF$_{01}$ &0.01  &0.04 &  0.49  &  0.76(2)  & 1.25 \\
DWF$_{02}$ &0.02  &0.04  &  0.48  &  0.80(3)  & 1.25 \\
DWF$_{03}$ &0.03  &0.04  &  0.61  &  0.66(2)  & 1.25 \\
\hline
MILC$_{f1}$ &0.006 &0.031  &  0.45 & 0.82(2) & 1.00 \\
MILC$_{f2}$ &0.012 &0.031  &  0.43 & 0.89(2) & 1.00 \\
\hline
\end{tabular}
\end{center}
\caption{The fitted parameters of mass function of DWF(RBC/UKQCD) and KS fermion (MILC).}
\end{table}

\section{The vector current quark-gluon vertex}
Near the cylinder cut and when the momentum transfer $q$ is small, the quark gluon vertex in momentum space is calculated from
\begin{equation}
\int d^4 x\int d^4 y e^{-ip(x-y)}G_{\mathcal O}(x,y)\nonumber\\
=\frac{1}{N}\sum_{i=1}^N \langle S_i(p|0)(\gamma_1 +\gamma_2 +\gamma_3)p (\gamma_5 S_i(p|0)^\dagger \gamma_5)\rangle \label{operator}
\end{equation}
where $S_i$ is a DWF propagator of the $i'$th sample among altogether $N$ samples.

\begin{figure}
\begin{minipage}[b]{0.47\linewidth}
\begin{center}
\includegraphics[width=.8\textwidth]{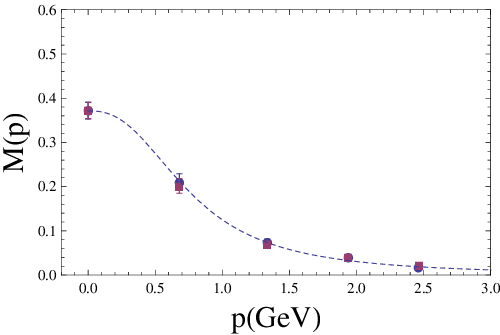}
\caption{The mass function of the domain wall fermion as a function of the modulus of Euclidean four momentum $p$.  DWF$_{01}$. (149 samples). Blue disks are $m_L$ (left handed quark) and red boxes are $m_R$ (right handed quark).}
\end{center}
\end{minipage}
\hfill
\begin{minipage}[b]{0.47\linewidth}
\begin{center}
\includegraphics[width=.8\textwidth]{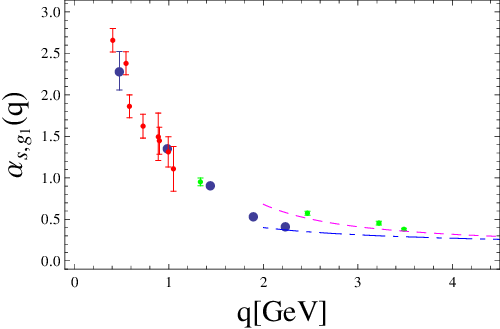}
\caption{The running coupling $\alpha_{s,g_1}(q)$ of MILC$_{f1}$ (blue disks) and DWF$_{01}$ (green points). The dash-dotted line is the pQCD result and the dashed line is the pQCD with the $\langle A^2\rangle$ condensate contribution. The red points are the experimental data of the JLab group.}\label{alpha}
\end{center}
\end{minipage}
\end{figure}

I specify the spin part of $\alpha\beta$ as $+-$ which corresponds to spin up for $\alpha$ and spin down for $\beta$. In the case of $\Gamma=q_4\gamma_i q_i$, 
I choose $\gamma_4{\gamma_1}^{+-}$ or $\gamma_4{\gamma_2}^{+-}$ as an example and ${\mathcal A}_i(p|0)^{-+}$ as the incoming wave ${\mathcal A}_i(p|0)^{++\,\dagger}$ contributes as the outgoing wave.
When ${\mathcal A}_i(p|0)^{--}$ is the incoming wave ${\mathcal A}_i(p|0)^{-+\,\dagger}$ contributes as the outgoing wave.@

The quark propagator is given as
\begin{equation}
S(p)=\frac{-i{\mathcal A}p+{\mathcal B}}{{\mathcal A} (p^2+{\mathcal{M M}}^\dagger)}
=\frac{ -i{\mathcal A} p+{\mathcal B}}{{\mathcal A} p^2+{\mathcal{M B}}^\dagger}
\end{equation}
I evaluate vector current matrix elements by diagonalizing the matrices in the
eq.(\ref{operator}) and getting the $\frac{1}{12}$ of the trace.

\subsection{The scalar form factor and the running coupling}
The vector current Ward identity allows us to extract the running coupling $\alpha_{s,g_1}(q)$ from the difference of $S^{-1}(p+\frac{\Vec q}{2})$ and $S^{-1}(p-\frac{\Vec q}{2})$ \cite{MPSTV94}
\begin{equation}
-i[S^{-1}((p+{\frac{\Vec q}{2}})_j|0)-S^{-1}((p-{\frac{\Vec q}{2}})_j|0)]=Z^V\Lambda_{0}(p)  q_j/4\pi.
\end{equation}
When a crossing is performed and a quark is transformed to anti-quark, $p$ can be treated as the momentum transfer $q$, to be compared with the experiment.
The green points above $q=1.4$ GeV in Fig.\ref{alpha} are the result of DWF. The error bars are taken from the Bootstrap method after 5000 re-samplings \cite{whit90,var}, which are smaller by about a factor of 10 as compared to the standard deviation of the bare samples. The data are comparable with the pQCD results with a phenomenological $\langle A^2\rangle$ condensates effect observed in the
ghost-gluon coupling of the MILC$_{f1}$ Landau gauge fixed configuration \cite{FN06}. Although statistics is not large, I observe that the running coupling $\alpha_{s,g_1}(q)$ of Landau gauge fixed configuration of MILC$_{f1}$ is not infrared suppressed in contrast to $\alpha_s(q)$ measured from the ghost-gluon vertex \cite{FN06}.

I compare the lattice results with experimental results extracted by the
JLab group \cite{DBCK06,DBCK08}. The JLab group analyzed the difference of the 
spin-dependent proton structure function and the spin-dependent neutron structure function as a function of $\displaystyle x=\frac{Q^2}{2M\nu}$ where $Q^2$
is the four-momentum squared, $\nu$ is the energy transfer and $M$ is the nucleon mass. The difference reduces the contribution of $\Delta(1232)$ and at
infinite momentum transfer squared the Bjorken sum rule \cite{Bj}, and at 0 momentum transfer the Gerasimov-Drell-Hearn sum rule \cite{Ger,DH} allows to fix the renormalization \cite{JLab08}. They extracted the running coupling in the infrared from
\begin{equation}
\Gamma_1^p-\Gamma_1^n=\frac{g_A}{6}[1-\frac{\alpha_s}{\pi}+o(\alpha_s^2)].
\end{equation}
In a confining theory where gluon have an effective mass, all vacuum polarization corrections to the gluon self-energy decouple at long wave length and 
one expects an infrared fixed point. 
A closeness of $\Gamma_1^p$ and $\Gamma_1^n$ implies a presence of infrared fixed point
$\alpha_{s,g_1}(0)=\pi$. When the coupling is constant in the infrared and the
quark masses can be ignored, one expects conformal symmetry or ADS/CFT correspondence to be applied in non-perturbative region \cite{BT06, Bro07}. 
Analytical Dyson-Schwinger approach also suggests the presence of infrared fixed point\cite{DS}.
\begin{figure}
\begin{minipage}[b]{0.47\linewidth}
\begin{center}
\includegraphics[width=.8\textwidth]{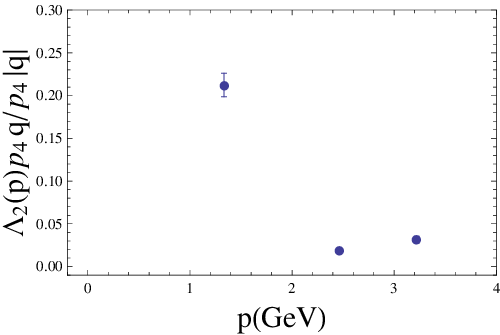}
\caption{The tensor form factor $\Lambda_2(p)p_4 \Vec q/p_4|\Vec q|$ at $|\Vec q|=2.6$GeV. (DWF$_{01}$, 52 samples)}\label{tensor}
\end{center}
\end{minipage}
\hfill
\begin{minipage}[b]{0.47\linewidth}
\begin{center}
\includegraphics[width=.8\textwidth]{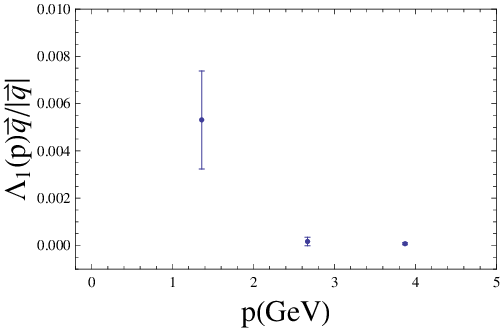}
\caption{The vector form factor $\Lambda_1(p) \Vec q/|\Vec q|$ at $|\Vec q|=2.6$GeV. (DWF$_{01}$, 52 samples)}\label{vector}
\end{center}
\end{minipage}
\end{figure}

\subsection{The tensor form factor}

I specify the spin of quarks that sandwich $\gamma_4\gamma_1, \gamma_4\gamma_2$ and $\gamma_4\gamma_3$ and pick up the quark propagator $S_i(p|0)$ whose
spin of the final state matches with that of $\gamma_4\gamma_i$. I assume
propagators are color diagonal. The vertices $\gamma_4\gamma_1$ and $\gamma_4\gamma_2$ are spin off-diagonal  and the vertex $\gamma_4\gamma_3$ is spin diagonal. There is a relative phase factor $i$ in $\gamma_4\gamma_2$ as compared to the other $\gamma_4\gamma_i$. 

The tensor term is evaluated from the difference of 
\begin{equation}
\langle\mathcal{A}_4 (p-\frac{q}{2}) \gamma_4\sum_j\gamma_j{\mathcal{A}_j}^\dagger(p+\frac{q}{2})\rangle p_4(p+\frac{q}{2})_j
-
\langle {\mathcal{A}_4}(p)\gamma_4\sum_j\gamma_j{\mathcal{A}_j}^\dagger(p)\rangle p_4 p_j
\end{equation}

I sample-wise diagonalize
\begin{eqnarray}
\Gamma_A^{L/R}=&&[\langle {\mathcal A^{L/R}}_4(p|0)p_4(\gamma_5 {\mathcal A}_i(p|0)^{\alpha\beta\,\dagger}\gamma_5)\sigma_1^{\alpha\beta}p_1\rangle\sigma_1\nonumber\\
&&+\langle{\mathcal A^{L/R}}_4(p|0)p_4(\gamma_5 {\mathcal A^{L/R}}_i(p|0)^{\alpha\beta\,\dagger}\gamma_5)\sigma_2^{\alpha\beta}p_2\rangle\sigma_2\nonumber\\
&&+\langle{\mathcal A^{L/R}}_4(p|0)(\gamma_5 {\mathcal A^{L/R}}_i(p|0)^{\alpha\beta\,\dagger}\gamma_5)\sigma_3^{\alpha\beta}p_3\rangle\sigma_3]\nonumber\\
\end{eqnarray} 
for each $\alpha, \beta$ that specify the color and the spin, 
and take the sum of the real part of the positive eigenvalues.
The corresponding momentum shifted matrix elements are
\begin{eqnarray}
\tilde\Gamma_A^{L/R}=&&[\langle {\mathcal A}_4^{L/R}(p-\frac{q}{2}|0)p_4(\gamma_5 {\mathcal A}_1^{L/R}(p+\frac{q}{2}|0)^{\alpha\beta\,\dagger}\gamma_5)\sigma_1^{\alpha\beta}(p+\frac{q}{2})_1\rangle\sigma_1\nonumber\\
&&+\langle{\mathcal A}_4^{L/R}(p-\frac{q}{2}|0)p_4(\gamma_5 {\mathcal A}_2^{L/R}(p+\frac{q}{2}|0)^{\alpha\beta\,\dagger}\gamma_5)\sigma_2^{\alpha\beta}(p+\frac{q}{2})_2\rangle\sigma_2\nonumber\\
&&+\langle{\mathcal A}_4^{L/R}(p-\frac{q}{2}|0)p_4(\gamma_5 {\mathcal A}_3^{L/R}(p+\frac{q}{2}|0)^{\alpha\beta\,\dagger}\gamma_5)\sigma_3^{\alpha\beta}(p+\frac{q}{2})_3\rangle\sigma_3].\nonumber
\end{eqnarray} 
I approximate the denominator of the propagator for $\Gamma_A$ and $\tilde \Gamma_A$ by those of the quark propagator in the cylinder cut 
and evaluate the tensor term as
\begin{equation}
\frac{1}{12}tr\frac{({\tilde\Gamma}_A^L+{\tilde\Gamma}_A^R )-({\Gamma_A}^L+{\Gamma_A}^R)}
{({\mathcal A} p^2+{\mathcal{M B}})({\mathcal A}^\dagger p^2+{\mathcal{M B}^\dagger})}=Z^V\Lambda_{2}(p) p_4 \Vec q_j/4\pi
\end{equation}
A simulation result of DWF$_{01}$ is shown in Fig.\ref{tensor}.

\subsection{The vector form factor}
I define the matrix elements between spin $++$ as $\Gamma_C$ and between spin $--$ as $\Gamma_D$.
\begin{eqnarray}
\Gamma_{C,j}&=&Re[\langle{\mathcal B}^{L,++}(p-\frac{\Vec q_j}{2}|0)\gamma_4(\gamma_5{\mathcal B}^{R,++}(p+\frac{\Vec q_j}{2}|0)^\dagger\gamma_5)\nonumber\\
&&-{\mathcal B}^{R,++}(p-\frac{\Vec q_j}{2}|0)\gamma_4(\gamma_5{\mathcal B}^{L,++}(p+\frac{\Vec q_j}{2}|0)^\dagger\gamma_5)\rangle]\nonumber
\end{eqnarray}
and
\begin{eqnarray}
\Gamma_{D,j}&=&Re[\langle{\mathcal B}^{L,--}(p-\frac{\Vec q_j}{2}|0)\gamma_4\gamma_5{\mathcal B}^{R,--}(p+\frac{\Vec q_j}{2}|0)^\dagger\gamma_5)\nonumber\\
&&-{\mathcal B}^{R,--}(p-\frac{\Vec q_j}{2}|0)\gamma_4(\gamma_5{\mathcal B}^{L,--}(p+\frac{\Vec q_j}{2}|0)^\dagger\gamma_5)\rangle]
\end{eqnarray}

\[
\frac{1}{6}\sum_j\frac{|\Gamma_{D,j}|+|\Gamma_{C,j}|}{({\mathcal A} p^2+{\mathcal{M B}})({\mathcal A}^\dagger p^2+{\mathcal{M B}^\dagger})}=Z^V\Lambda_{1}(p)  q_j/4\pi
\]
A simulation result of the vector form factor of DWF$_{01}$ is shown in Fig.\ref{vector}.

\section{Discussion}
I showed that nonperturbative renormalization of lattice Coulomb gauge quark-gluon vertices is feasible. In addition to the running coupling, I measured the tensor term $g_3(p,\Vec q)\gamma_4 p_4\Slash{q}$ and the vector term $g_2(p,\Vec q)\Vec q$ of the Coulomb gauge quark-gluon vertex. 

The running coupling $\alpha_I(q)$ and $\alpha_{s,g_1}(q)$ of DWF are consistent with the JLab extraction. The running coupling of Landau gauge $\alpha_{s,g_1}(q)$ is not infrared suppressed in contrast to $\alpha_s(q)$ of triple gluon vertex \cite{Orsay03} and ghost-glion vertex \cite{FN07,FN06}. The infrared suppression of $\alpha_s(q)$ in Landau gauge was attributed to an instanton effect \cite{Orsay03}. However, as shown by 't Hooft \cite{tH86}, in the instanton calculus extended to the infrared region, there is a divergence from zero modes. In supersymmetric case, however, the divergence from fermionic zero mode and bosonic zero mode cancel \cite{AdVe77} and a finite infrared fixed point appears. It is plausible that the search of a stable solution in Coulomb gauge with use of the conjugate gradient method leads to a solution in which zero mode divergences of gluons and those of quarks cancel.

When DWF configurations of larger lattices are available, the $\alpha_{s,g_1}(q)$ 
data can be extended to lower energies. I expect that $\alpha_{s,g_1}(q)$ of DWF in Coulomb gauge show the same behavior as $\alpha_I(q)$.
The difference of the infrared features of Landau gauge and Coulomb gauge
running coupling casts questions on the Kugo-Ojima confinement criterion \cite{SF07,SF08a} which is derived in Landau gauge but effects of instantons are not 
considered. 
Simulations with momenta far from the cylinder cut and a calculation of the form factor of three quark systems are also left in the future. 

\leftline{\bf Acknowledgement}

I thank J.P. Chen and A. Deur for the discussion on the extraction of
 running coupling from the experimental data, R. Alkofer for the
inquiry on the tensor form factor which became a motivation of the 
present work, and H. Nakajima for the help in the gauge fixing.

The numerical simulation was performed on Hitachi-SR11000 at High Energy Accelerator Research Organization(KEK) under a support of its Large Scale Simulation Program (No.07-04 and No.08-01), and on NEC-SX8 at Yukawa institute of theoretical physics of Kyoto University.

\end{document}